# The role of artificial intelligence in achieving the Sustainable Development Goals

*The emergence of artificial intelligence (AI) and its progressively wider impact on many sectors across the society requires an assessment of its effect on sustainable development. Here we analyze published evidence of positive or negative impacts of AI on the achievement of each of the 17 goals and 169 targets of the 2030 Agenda for Sustainable Development. We find that AI can support the achievement of 128 targets across all SDGs, but it may also inhibit 58 targets. Notably, AI enables new technologies that improve efficiency and productivity, but it may also lead to increased inequalities among and within countries, thus hindering the achievement of the 2030 Agenda. The fast development of AI needs to be supported by appropriate policy and regulation. Otherwise, it would lead to gaps in transparency, accountability, safety and ethical standards of AI-based technology, which could be detrimental towards the development and sustainable use of AI. Finally, there is a lack of research assessing the medium- and long-term impacts of AI. It is therefore essential to reinforce the global debate regarding the use of AI and to develop the necessary regulatory insight and oversight for AI-based technologies.*

**Ricardo Vinuesa[1]\*, Hossein Azizpour[2], Iolanda Leite[3], Madeline Balaam[4], Virginia Dignum[5], Sami Domisch[6], Anna Felländer[7], Simone Langhans[8,9], Max Tegmark[10] and Francesco Fuso Nerini[11]\*\***

[1] Linné FLOW Centre, KTH Mechanics, SE-100 44 Stockholm, Sweden.
[2] School of Electrical Engineering and Computer Science, KTH Royal Institute Of Technology, Stockholm, Sweden
[3] Division of Robotics, Perception, and Learning, School of EECS, KTH Royal Institute Of Technology, Stockholm, Sweden
[4] Division of Media Technology and Interaction Design, KTH Royal Institute of Technology, Lindstedtsvägen 3, Stockholm, Sweden
[5] Responsible AI Lab, Umeå University, SE-90358 Umeå, Sweden.
[6] Leibniz-Institute of Freshwater Ecology and Inland Fisheries, Müggelseedamm 310, 12587 Berlin, Germany.
[7] Division of Media Technology and Interaction Design, KTH Royal Institute of Technology, Lindstedtsvägen 3, Stockholm, Sweden
[8] Basque Centre for Climate Change (BC3), Leioa, Spain.
[9] Department of Zoology, University of Otago, 340 Great King Street, 9016 Dunedin, New Zealand.
[10] Center for Brains, Minds & Machines, Massachusetts Institute of Technology, Cambridge, MA 02139, USA
[11] Unit of Energy Systems Analysis (dESA), KTH Royal Institute of Technology, Brinellvagen 68, SE-100 44. Stockholm, Sweden.

\*rvinuesa@mech.kth.se;\*\*francesco.fusonerini@energy.kth.se

The emergence of artificial intelligence (AI) is shaping an increasing range of sectors. For instance AI is expected to affect global productivity[3], equality and inclusion[4], environmental outcomes[5] and several other areas, both in the short and long term[6]. Reported potential impacts of AI indicate both positive[7] and negative[8] impacts on sustainable development. However, to date there is no published study systematically assessing the extent to which AI might impact all aspects of sustainable development – defined in this study as the 17 interconnected Sustainable Development Goals (SDGs) and 169 targets internationally agreed in the 2030 Agenda for Sustainable Development[1]. This is a critical research gap, since we find that AI may influence the ability to meet all Sustainable Development Goals (see a summary of the results in Fig. 1, and full results in the Supplementary Table 1).

Here we present and discuss implications of how AI can either enable or inhibit the delivery of all 17 Goals and 169 targets recognized in the 2030 Agenda for Sustainable Development. Relationships were characterized by the methods reported at the end of this article, which can be summarized as a consensus-based expert elicitation process, informed by previous studies aimed at mapping SDGs interlinkages[9]. For this study, we adopt Russell and Norvig's definition of AI as a field that "attempts not just to understand but to build intelligent entities"[2] (see full definition in the Methods section). This view encompasses a large variety of subfields, including machine learning.

**Documented connections between AI and the SDGs**

Our review of relevant evidence shows that AI may act as an enabler on 128 targets (76%) across all SDGs, generally through a technological improvement which may allow to overcome certain present limitations. However, 58 targets (34%, also across all SDGs) may experience a negative impact from the development of AI. For the purpose of this study, we divide the SDGs into three categories, according to the three pillars of sustainable development, namely *Society*, *Economy* and *Environment* [10,11] (see the Methods section). This classification allows us to provide an overview of the general areas of influence of AI.

**Fig. 1 Summary of positive and negative impact of AI on the various SDGs.** Documented evidence of the potential of AI acting as **a** an enabler or **b** an inhibitor on each of the SDGs. The percentages in the inner circle of the figure indicate the proportion of targets within an SDG potentially affected by AI. The results corresponding to the three main groups, namely *Society*, *Economy* and *Environment*, are also shown in the outer circle of the figure.

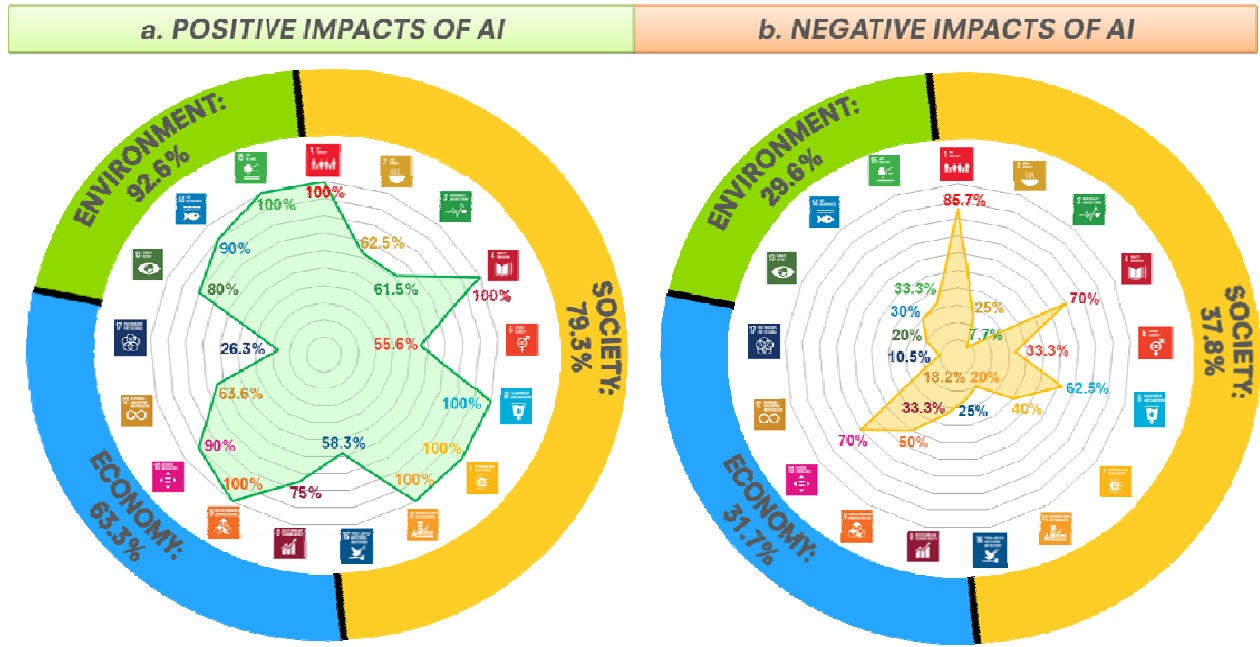

A detailed assessment of the *Society*, *Economy* and *Environment* groups, together with illustrative examples, are discussed next.

**AI and societal outcomes.** 65 targets (79%) within the *Society* group could potentially benefit from AI-based technologies (Fig. 2). For instance in SDG 1 on no poverty, SDG 4 on quality education, SDG 6 on clean water and sanitation, SDG 7 on affordable and clean energy and SDG 11 on sustainable cities AI may act as an enabler for all the targets by providing food, health, water and energy services to the population. It can also support low-carbon systems, for instance by supporting the creation of smart and circular cities that efficiently use their resources[12]. Fewer targets in the *Society* group can be impacted negatively by AI (31 targets, 38%) than the ones with positive impact. But their consideration is crucial. Many of these relate to how the technological improvements enabled by AI may be implemented in countries with different cultural values and wealth. Advanced AI technology, research and product design can require massive computational resources only available through large computing centers. These facilities have a very high energy requirement and carbon footprint[13]. For instance, new applications such as bitcoin globally are using as much electricity as some nations´ electrical demand[14], compromising outcomes in the SDG 7 sphere, but also on SDG 13 on Climate Action. Although AI-enabled technology can act as a catalyst[7] to achieve the 2030 Agenda, it may also trigger inequalities which may act as inhibitors on SDGs 1, 4 and 5. This is reflected for instance in target 1.1, since AI can help to identify areas of poverty and foster international action using satellite images[7]. On the other hand, it may also lead to additional qualification requirements for any job, consequently increasing the inherent inequalities[15] and acting as an inhibitor towards the achievement of this target.

Another important drawback of AI-based developments is that they are traditionally based on the needs and values of nations in which AI is being developed. For instance, complex AI-enhanced agricultural equipment may not be accessible to small farmers and thus produce an increased gap with respect to larger producers in more developed economies[16], consequently inhibiting the achievement of some targets of SDG 2 on Zero Hunger. There is another important shortcoming of AI in the context of SDG5 on gender equality: there is insufficient research assessing the potential impact of technologies such as smart algorithms, image recognition or reinforced learning on discrimination against women and minorities. For instance, the fact that voice agents and chatbots (such as Alexa or Siri) are gendered as female may continue to reinforce stereotypes of women as personal assistants to men[4].

**Fig. 2 Detailed assessment of the impact of AI on the SDGs within the *Society* group.** Documented evidence of positive or negative impact of AI on the achievement of each of the targets from SDGs 1, 2, 3, 4, 5, 6, 7, 11 and 16. Each block in the diagram represents a target. For targets highlighted in green or orange we found published evidence that AI could potentially enable or inhibit such target, respectively. The absence of highlighting indicates absence of identified evidence. Note that this does not necessarily imply the absence of a relationship.

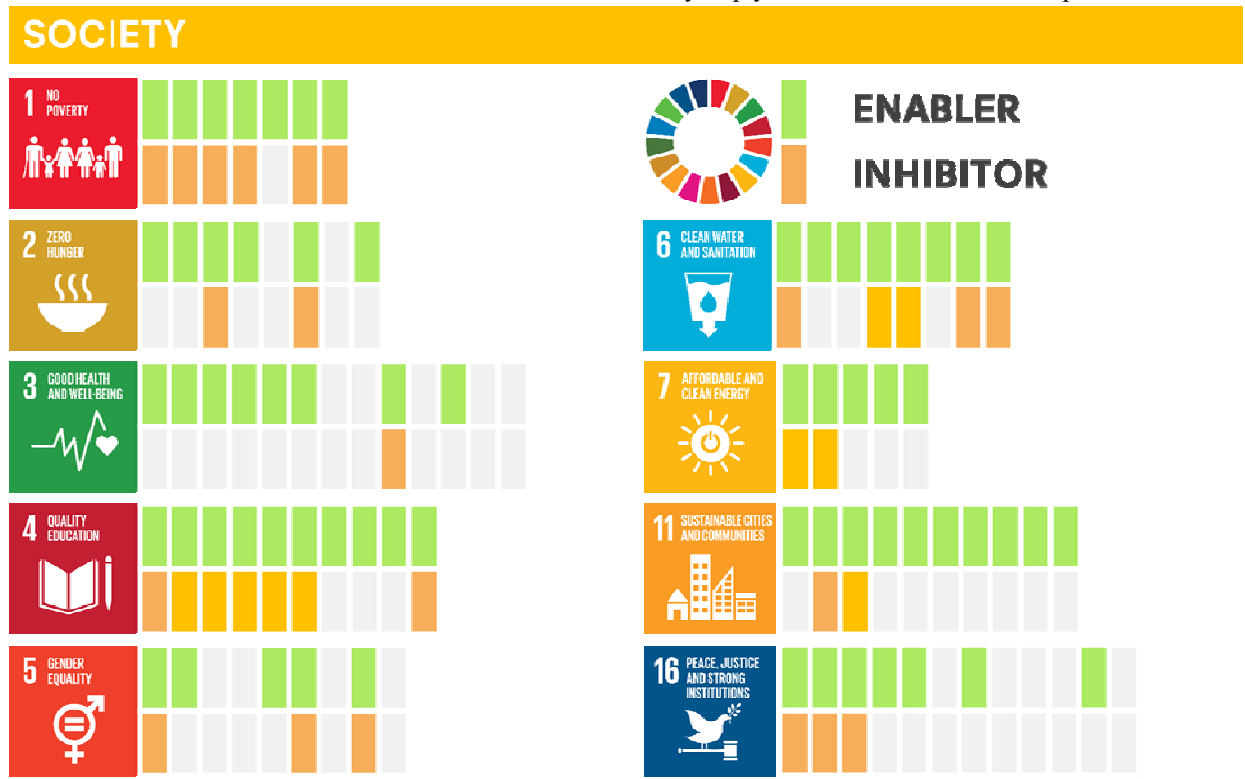

**AI and economic outcomes.** The technological advantages provided by AI may also have a positive impact on the achievement of a number of SDGs within the *Economy* group. We have identified benefits from AI on 38 targets(63%) from these SDGs, whereas negative impacts are reported in 19 targets(32%) (Fig. 1). Although Acemoglu and Restrepo (2018)[3] report a net positive impact of AI-enabled technologies[17–21], associated to increased productivity, the literature also reflects potential negative impacts mainly associated to increased inequalities. In the context of the *Economy* group of SDGs, if future markets rely heavily on data analysis, and these resources are not equally available in low- and middle- income countries, the economical gap may be significantly increased due to the newly introduced inequalities[22,23] significantly impacting SDGs 8 (decent work and economic growth), 9 (industry, innovation and infrastructure) and 10 (reduced inequalities). Erik Brynjolfsson and Andrew McAfee (2014)[24] argue that AI can exacerbate inequality also within nations. By replacing old jobs with ones requiring more skills, technology disproportionately rewards the educated: since the mid 1970s, US salaries rose about 25% for those with graduate degrees while the average high school dropout took a 30% pay cut. Moreover, automation shifts corporate income to those who own companies from those who work there. Such transfer of revenue from workers to investors helps explain why, even though the combined revenues of Detroit's "Big 3" (GM, Ford and Chrysler) in 1990 were almost identical to those of Silicon Valley's "Big 3" (Google, Apple, Facebook) in 2014, the latter had nine times fewer employees and were worth thirty times more on the stock market[25]. Fig. 3 shows an assessment of the documented positive and negative effects on the various targets within the SDGs in the *Economy* group.

While the found linkages in the *Economy* group are mainly positive, trade-offs cannot be neglected. For instance, AI can have a negative effect on social media usage, by showing users content specifically suited to their preconceived ideas. This may lead to political polarization[26], with consequences in the context of SDG10 on reduced inequalities. On the other hand, AI can help identify sources of inequality and conflict[27,28], and therewith potentially reduce inequalities, for instance by using simulations to assess how virtual societies may respond. But there is an underlying risk when using AI to evaluate and predict human behavior, which is the inherent bias in the data. It has been reported that a number of discriminatory challenges are faced in the automated targeting of online job advertising using AI[28], essentially related to the previous biases in selection processes conducted by human recruiters. The work by Dalenberg (2018)[28] highlights the need of modifying the data preparation process and explicitly adapting the AI-based algorithms used for selection processes in order to avoid such biases.

**Fig. 3 Detailed assessment of the impact of AI on the SDGs within the *Economy* group.** Documented evidence of positive or negative effect of AI on the achievement of each of the targets from SDGs 8, 9, 10, 12 and 17. The interpretation of the blocks and colors is as in Fig. 2.

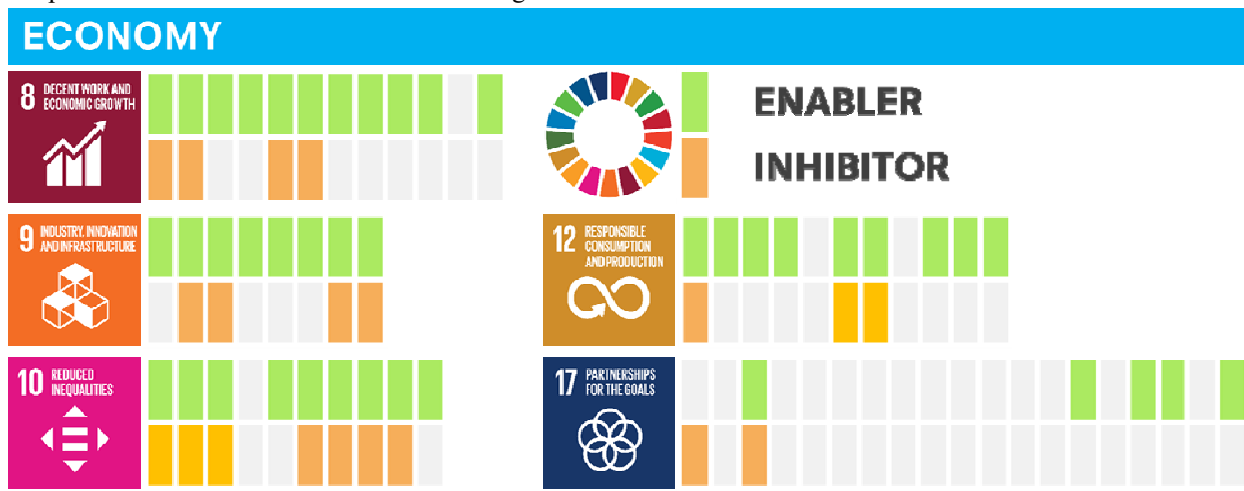

**AI and environmental outcomes.** The last group of SDGs, i.e. the one related to *Environment*, is analyzed in Fig. 4. The three SDGs in this group are related to climate action, life below water and life on land (SDGs 13, 14 and 15). For the *Environment* group, we identified 25targets(93%)for which AI could act as an enabler. Benefits from AI could derive by the possibility of analyzing large-scale interconnected databases to develop joint actions aimed at preserving the environment. Looking at SDG 13 on Climate Action, there is evidence that AI advances will support the understanding of climate change, and the modelling of its possible impacts. Further, AI will support low-carbon energy systems with high integration of renewable energy and energy efficiency, which are all needed to address climate change[29].Broader ecosystems could also benefit. The achievement of target 14.1, calling to prevent and significantly reduce marine pollution of all kinds, can benefit from AI through algorithms for automatic identification of possible oil spills[30]. Another example is target 15.3 which calls for combating desertification and restoring degraded land and soil. According to Mohamadi et al. (2016)[31], neural networks and objective-oriented techniques can be used to improve the classification of vegetation cover types based on satellite images, with the possibility of processing large amounts of images in a relatively short time. These AI techniques can help to identify desertification trends over large areas, information that is relevant for environmental planning, decision-making and management to avoid further desertification, or help reverse trends by identifying the major drivers. However, we also found that such increased access to AI-related information of ecosystems may drive over-exploitation of resources[32].

**Fig. 4 Detailed assessment of the impact of AI on the SDGs within the *Environment* group.** Documented evidence of positive or negative impact of AI on the achievement of each of the targets from SDGs 13, 14 and 15. The interpretation of the blocks and colors is as in Fig. 2.

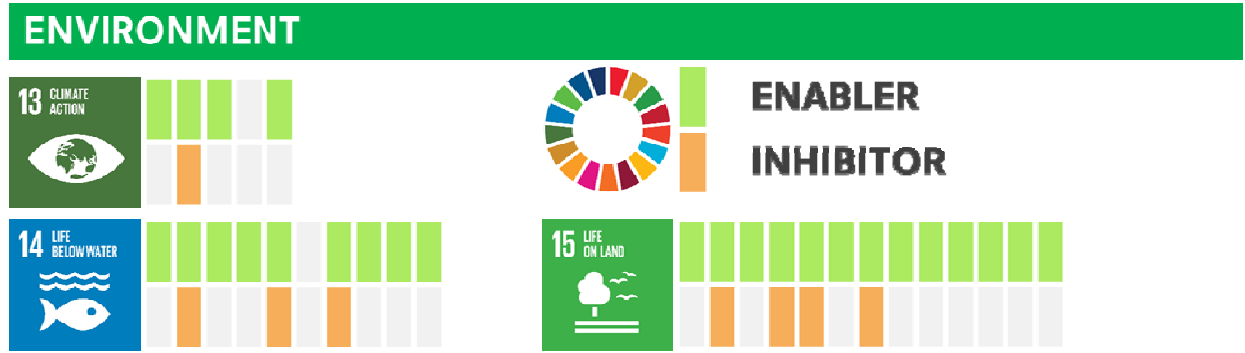

**Research Gaps on the role of AI in Sustainable Development**
The more we enable SDG's by AI-deployment, from autonomous vehicles toAI-powered healthcare solutions[33] and smart electrical grids[12], the more important it becomes to invest in the AI safety research needed to keep these systems robust and beneficial, to prevent them from malfunctioning or getting hacked[34]. Also, while we were able to find numerous studies suggesting that AI can potentially serve as an enabler for many SDG targets and indicators, a significant fraction of these studies have been conducted in controlled laboratory environments, based on limited datasets or using prototypes[35–37]. Hence, extrapolating this information to evaluate the real-world effects often remains a challenge. This is particularly true when measuring the impact of AI across broader scales, both temporally and spatially. We acknowledge that conducting controlled experimental trials for evaluating real-world impacts of AI can result in depicting a snapshot situation, where AI tools are tailored towards that specific environment. However, as society is constantly changing (also due to factors including non-AI-based technological advances), the requirements set for AI are changing as well, resulting in a feedback loop with interactions between society and AI. Therefore, novel methodologies are required to ensure that the impact of new technologies are assessed from the points of view of efficiency, ethics and sustainability, prior to launching large-scale AI deployments.

Although we found more published evidence of AI serving as an enabler than as an inhibitor, there are at least two important aspects that should be considered. First, self-interest should be expected to bias the AI research community and industry towards publishing positive results. Second, discovering detrimental aspects of AI may require longer longitudinal studies and, as mentioned above, there are not many established evaluation methodologies available to do so. Bias towards publishing positive results is particularly apparent in the SDGs corresponding to the *Environment* group. A good example of this bias is regarding target 14.5 on conserving coastal and marine areas, where machine-learning algorithms can provide optimum solutions given a wide range of parameters regarding the best choice of areas to include in conservation networks[38]. However, even if the solutions are optimal from the mathematical point of view (given a certain range of selected parameters), additional research would be needed in order to assess the long-term impact of such algorithms regarding equity and fairness[8], precisely because of the unknown factors that may come into play. Regarding the second point stated above, it is likely that the AI projects with highest potential of maximizing profit will get funded. This likely results in increased inequality[39]. Consequently, there is the risk that AI-based technologies with potential to achieve certain

SDGs may not be prioritized, if their expected economic impact is not high. Furthermore, it is essential to promote the development of initiatives to assess the ethical implications of new AI technologies.

Substantive research and application of AI technologies to SDGs is concerned with measuring or predicting certain events using, for example, data mining and machine-learning techniques. This is the case of applications such as forecasting extreme weather events or predicting recidivist offender behavior. The expectation with this research is to allow the preparation and response for a wide range of events. However, there is a research gap in real-world applications of such systems, e.g. by governments (as discussed above). Barriers for institutions for adopting AI systems and data as part of their decision-making process include the possibility to adopt such technology while protecting the privacy of citizens and data, high cyber-security needs, and the technical capabilities needed to have AI-systems functioning properly. Targeting these gaps would be essential to ensure the usability and practicality of AI technologies for governments. This would also be a prerequisite for understanding long-term impacts of AI regarding its potential, while regulating its use to reduce the possible bias that can be inherent to AI[8].

Furthermore, our research suggests that AI applications are currently biased towards SDG issues that are mainly relevant to those nations where most AI researchers live and work. For instance, many systems applying AI technologies to agriculture, e.g. to automate harvesting or optimize its timing, are located within wealthy Western nations. Our literature search resulted in only a handful of examples where AI technologies are applied to SDGs-related issues in nations without strong AI research. Moreover, if AI technologies are designed and developed for technologically advanced settings, they have the potential to exacerbate problems in less wealthy nations (e.g. when it comes to food production). This finding leads to a substantial concern that developments in AI technologies could increase inequalities between wealthy and less wealthy nations, in ways which counteract the overall purpose of the SDGs. We encourage researchers and funders to focus more on designing and developing AI solutions which respond to localized problems in less wealthy nations. Projects undertaking such work should ensure that solutions are not simply transferred from a wealthy nation. Instead, they should be developed based on a deep understanding of the respective region or culture to increase the likelihood of adoption and success.

**Towards Sustainable AI**
Our assessment of published evidence shows that AI can have a positive impact on all the SDGs. This is essentially through technological breakthroughs that will lead to better outcomes in several sectors. However, there are a number of problems associated with AI that if not addressed may inhibit the achievement of several SDGs.

First, the great wealth that AI-powered technology has the potential to create may go mainly to those already well-off and educated, while job displacement leaves others worse off. Globally, the growing economic importance of AI may result in increased inequalities due to the unevenly distributed educational and computing resources throughout the world. Furthermore, the existing biases in the data used to train AI algorithms may result in the exacerbation of those biases, eventually leading to increased discrimination. Other related problems are the political polarization due to the massive use of social media, the lack of robust research methods to assess the long-term impact of AI, and privacy issues related to the data-intensiveness of AI applications. Many of these aspects result from the interplay between technology development, requests from individuals and response from governments. Figure 5 shows a schematic representation of these dynamics.

**Fig. 5 Interaction of AI and society.** Schematic representation showing the identified agents and their roles towards the development of AI. Thicker arrows indicate faster speed of change. In this representation, technology affects individuals through technical developments, which change the way people work and interact with each other and with the environment, whereas individuals would interact with technology through new needs to be satisfied. Technology (including technology itself and its developers) affects governments through new developments that need appropriate piloting and testing. Also, technology developers affect government through lobbying and influencing decision makers. Governments provide standards and regulations to technology. Finally, the governments affect individuals through policy and legislation, and individuals would require from the government new regulations consistent with the changing circumstances. The environment is an underlying layer that gives the "planetary boundaries" to the mentioned interactions.

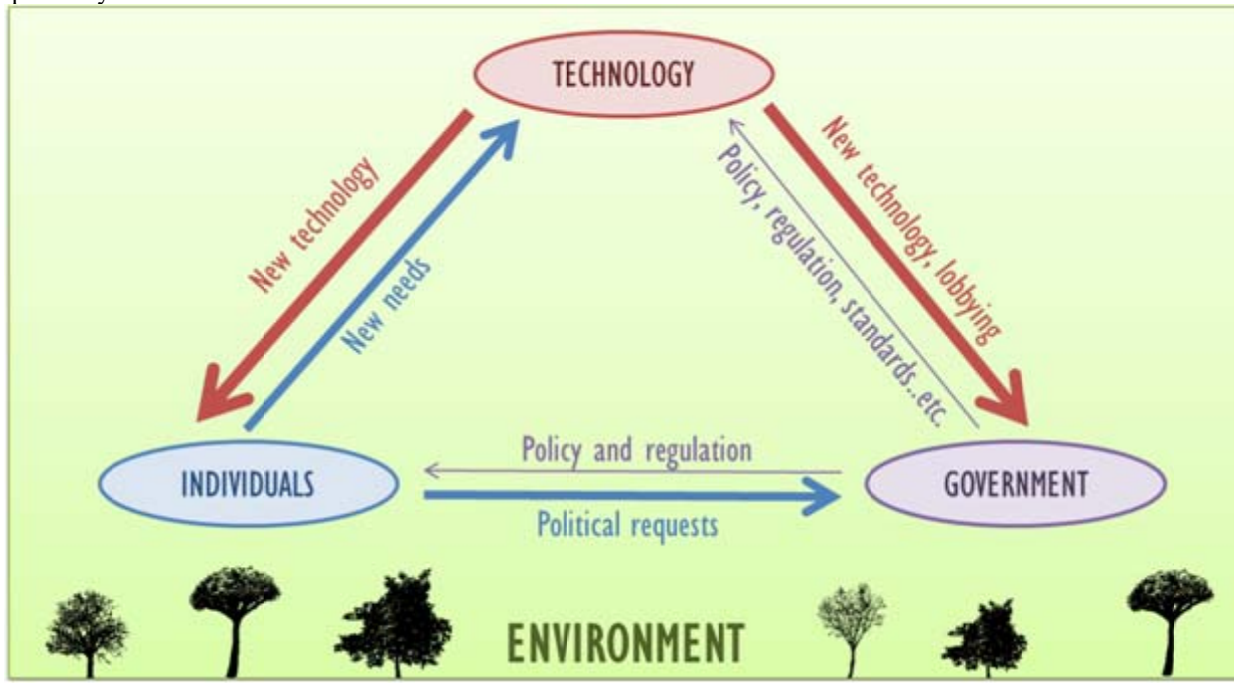

Based on the evidence discussed above, these interactions are not currently balanced, and the advent of AI has exacerbated the process. A wide range of new technologies are being developed very fast, significantly affecting the way individuals live, requiring new piloting procedures from governments. The problem is that neither individuals nor governments seem able to follow the pace of these technological developments. This fact is illustrated by the lack of appropriate regulations to ensure the long-term viability of these new technologies. We argue that it is essential to reverse this trend. A first step in this direction is to establish adequate policy and regulation frameworks, to help direct the vast potential of AI towards the highest benefit for individuals and the environment as well as for the achievement of the SDGs. Regulatory oversight should be preceded by regulatory *insight*, where policymakers have sufficient understanding of AI challenges to be able to formulate sound policy. Developing such insight is even more urgent than oversight, since policy formulated without understanding is likely to be ineffective at best and counterproductive at worst.

While strong and connected institutions (covered by SDG 16) are needed to regulate the future of AI, we find that there is limited understanding of the potential impact of AI on institutions. Examples of the positive impacts include AI algorithms aimed at improving fraud detection[40,41] or assessing the possible effects of certain regulations[42,43]. On the other hand, the data intensity of AI-approaches limits their uptake by institutions due to privacy and security concerns. Another concern is that data-driven approaches for policing may hinder equal access to justice because of algorithm bias, particularly towards minorities[44].

It is also imperative to develop regulations regarding transparency and accountability of AI, as well as to decide the ethical standards to which AI-based technology should be subjected to. This debate is being pushed forward by initiatives such as the IEEE ethical aligned design[45], and the new EU ethical guidelines for trustworthy AI[46]. In this sense, the lack of interpretability of AI, which is currently one of the challenges of AI research, adds an additional complication to the enforcement of such regulatory actions[47]. This, however, implies that AI algorithms, which are trained with data consisting of previous regulations and decisions, may act as a "mirror" reflecting biases and unfair policy. This presents an opportunity to possibly identify and correct certain errors in the existing procedures. Again, the friction between the uptake of data-driven AI applications and the need of protecting individuals´ privacy and security is stark. When not properly regulated, the vast amount of data produced by the citizens might potentially be used to influence consumer opinion towards a certain product or political cause[48].

We are at a critical turning point for the future of AI. A global and science-driven debate to develop shared principles and regulations among nations and cultures is necessary to shape a future in which AI positively contributes to the achievement of all the Sustainable Development Goals. All actors in all nations should be represented in this dialogue, to ensure that no one is left behind. On the other hand, postponing or not having such conversation could result in an unequal and unsustainable AI-fueled development.

**Methods**

In this section we describe the process employed to obtain the results described in the present study and shown in the Supplementary Table 1. The goal was to answer the question "Is there published evidence of AI acting as an enabler or an inhibitor for this particular target?", for each of the 169 targets within the 17 SDGs. To this end, we conducted a consensus-based expert elicitation process, as discussed by Butler et al. (2015)[49] and Morgan (2014)[50]. The authors of this paper are academics spanning a wide range of disciplines, including engineering, natural and social sciences, and acted as experts for the elicitation process. The authors performed an expert-driven literature search to support the identified connections between AI and the various targets, where the following sources of information were considered as acceptable evidence:
- Published evidence on real-world applications (given the quality variation depending on the venue, we ensured that the publications considered in the analysis were of sufficient quality).
- Published evidence on controlled/laboratory scenarios (given the quality variation depending on the venue, we ensured that the publications considered in the analysis were of sufficient quality).
- Reports from accredited organizations (for instance: UN or government bodies).
- Documented commercial-stage applications.

On the other hand, the following sources of information were not considered as acceptable evidence:
- Educated conjectures, real-world applications without peer-reviewed research.
- Media, public beliefs, etc.
- Other sources of information.

We considered any software technology with at least one of the following capabilities as relevant: *perception* – including audio, visual, textual, and tactile (e.g. face recognition), *decision-making* (e.g. medical diagnosis systems), *prediction* (e.g. weather forecast), *automatic knowledge extraction and pattern recognition from data* (e.g. discovery of fake news circles in social media), *interactive communication* (e.g. social robots or chat bots) and *logical reasoning* (e.g. theory development from premises). The list of connections between AI and the 169 targets, together with a paragraph summarizing the reasoning behind the assessment of AI acting as a potential enabler or inhibitor of that particular target, is available in the Supplementary Table 1. A list of references supporting the reasoning is also provided for each of the targets.

The expert elicitation process was conducted as follows: each of the SDGs was assigned to one or more main contributors, and in some cases to several additional contributors as summarized in Table 1. The main contributors carried out a first literature search for that SDG, and assessed whether the published evidence reflected positive or negative impacts of AI on the various targets of that particular SDG. Then the additional contributors (if assigned to that SDG) completed the main analysis with additional references and discussions with the main contributors. One published study on a synergy or a trade-off between a target and AI was considered enough for mapping the interlinkage. However, for nearly all targets several references are provided. After the analysis of a certain SDG was concluded by the contributors, a reviewer was assigned to evaluate the connections and reasoning presented by the contributors. The reviewer was not part of the first analysis, and we tried to assign the roles of main contributor and reviewer to experts with complementary competences for each of the SDGs. The role of the reviewer was to bring up additional points of view and considerations, while critically assessing the analysis. Then main contributors and reviewers iteratively discussed to improve the results presented for each of the SDGs. This process was conducted through regular meetings over approximately six months, until the analysis for all the SDGs was sufficiently refined.

After reaching consensus regarding the assessment shown in Supplementary Table 1, we analyzed the results quantitatively by evaluating the number of targets for which AI may act as an enabler or an inhibitor. A total of 128 targets reflected positive impact of AI, whereas for 58 the literature indicated negative impact. This corresponds to 75.7% and 34.3% of targets with positive and negative impact, respectively. Furthermore, we carried out the same analysis for each of the SDGs, and calculated the percentage of targets with positive and negative impact of AI for each of the 17 Goals, as shown in Fig. 1. Additionally, we divided the SDGs into the three following categories: *Society, Economy* and *Environment,* consistent with the classification discussed by Refs.[10,11]. The SDGs assigned to each of the categories are shown in Fig. 6 and the individual results from each of these groups can be observed in Figs. 2-4. These figures indicate, for each target within each SDG, whether any published evidence of positive or negative impact was found.

**Fig. 6 Categorization of the SDGs into the *Economy, Society* and *Environment* groups.**

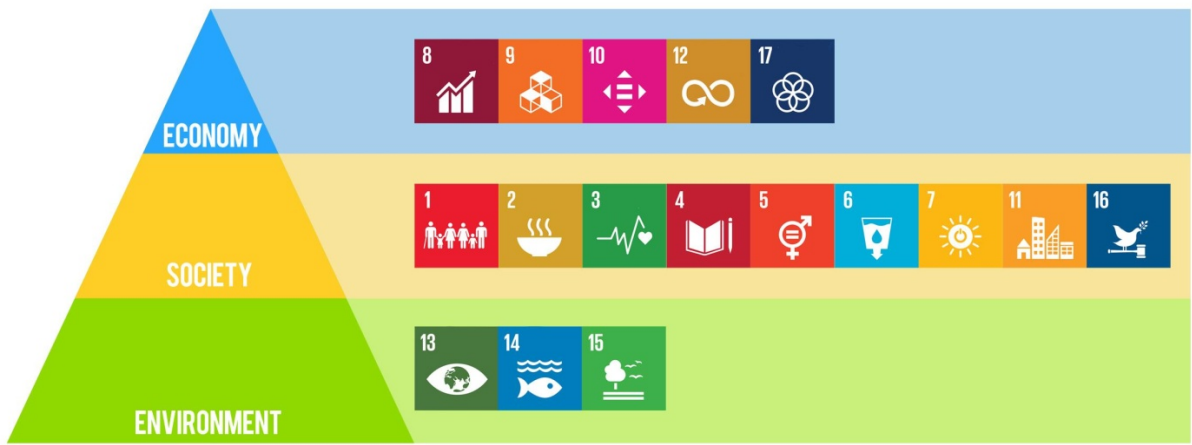

**Table 1 Experts assigned to each SDG.** Summary of the roles from the various experts in the analysis of the various SDGs. The initials correspond to the author names.

| SDG | Main contributors | Additional contributors | Reviewer |
| --- | --- | --- | --- |
| 1: No poverty | RV | | MB |
| 2: Zero hunger | MB | | FFN |
| 3: Good health and well-being | HA | SL | IL |
| 4: Quality education | RV | IL | MB |
| 5: Gender equality | MB | | RV |
| 6: Clean water and sanitation | SL | SD | FFN |
| 7: Affordable and clean energy | FFN | | RV |
| 8: Decent work and economic growth | AF, FFN | HA | SD |
| 9: Industry, innovation, and infrastructure | RV | | IL |
| 10: Reducing inequalities | SL, RV | AF, MB, SL | FFN |
| 11: Sustainable cities and communities | RV | | HA |
| 12: Responsible consumption and production | AF, FFN, RV | | SL |
| 13: Climate action | FFN | | HA |
| 14: Life below water | SD | SL | RV |
| 15: Life on land | SL | SD | RV |
| 16: Peace, justice and strong institutions | IL | | AF |
| 17: Partnerships for the Goals | AF, HA, MB | | SL, SD, RV |

**Authors Contributions**

R.V. and F.F.N. ideated, designed and wrote the paper; they also coordinated inputs from the other authors, assessed and reviewed SDG evaluations as for Table 1. H.A. and I.L. supported the design, wrote and reviewed sections of the paper; they also assessed and reviewed SDG evaluations as for Table 1. M.B., S.D., A.F. and S.L. wrote and reviewed sections of the paper; they also assessed and reviewed SDG evaluations as for Table 1. M.T. reviewed the paper and acted as final editor. V.D. reviewed sections of the paper.

## References

1. UN General Assembly (UNGA). Transforming our world: the 2030 Agenda for Sustainable Development. *Resolut. A/RES/70/1, 25* 1–35 (2015).

2. Russell, S. & Norvig, P. *Artificial Intelligence: A Modern Approach, Global Edition*. (2016).

3. Acemoglu, D. & Restrepo, P. *Artificial Intelligence, Automation and Work*. (2018). doi:10.3386/w24196

4. Søndergaard, M. L. J. & Hansen, L. K. Intimate Futures: Staying with the Trouble of Digital Personal Assistants Through Design Fiction. *Proceedings of the 2018 Designing Interactive Systems Conference* 869–880 (2018). doi:10.1145/3196709.3196766

5. Norouzzadeh, M. S. *et al.* Automatically identifying, counting, and describing wild animals in camera-trap images with deep learning. *Proc. Natl. Acad. Sci. U. S. A.* **115,** E5716–E5725 (2018).

6. Tegmark, M. *Life 3.0 : being human in the age of artificial intelligence*. (2017).

7. Jean, N. *et al.* Combining satellite imagery and machine learning to predict poverty. *Science (80-. ).* **353,** 790–794 (2016).

8. Courtland, R. Bias detectives: the researchers striving to make algorithms fair. *Nature* **558,** 357–360 (2018).

9. Fuso Nerini, F. *et al.* Mapping synergies and trade-offs between energy and the Sustainable Development Goals. *Nat. Energy* (2017). doi:10.1038/s41560-017-0036-5

10. UN. Sustainable Development | UNITED NATIONS ECONOMIC and SOCIAL COUNCIL. (2019). Available at: https://www.un.org/ecosoc/en/sustainable-development. (Accessed: 23rd January 2019)

11. *Stockholm Resilience Centre's (SRC) contribution to the 2016 Swedish 2030 Agenda HLPF report*. (2017).

12. International Energy Agency. *Digitalization & Energy*. (2017).

13. Brown, R. & et al. *Report to Congress on Server and Data Center Energy Efficiency: Public Law 109-431 | Energy Technologies Area*. (2008).

14. Truby, J. Decarbonizing Bitcoin: Law and policy choices for reducing the energy consumption of Blockchain technologies and digital currencies. *Energy Res. Soc. Sci.* **44,** 399–410 (2018).

15. Nagano, A. Economic Growth and Automation Risks in Developing Countries Due to the Transition Toward Digital Modernity. *Proceedings of the 11th International Conference on Theory and Practice of Electronic Governance - ICEGOV '18* (2018). doi:10.1145/3209415.3209442

16. Wegren, S. K. The "left behind": Smallholders in contemporary Russian agriculture. *J. Agrar. Chang.* **18,** 913–925 (2018).

17. Cockburn, I., Henderson, R. & Stern, S. *The Impact of Artificial Intelligence on Innovation*. (2018). doi:10.3386/w24449

18. Seo, Y., Kim, S., Kisi, O. & Singh, V. P. Daily water level forecasting using wavelet decomposition and artificial intelligence techniques. *J. Hydrol.* **520,** 224–243 (2015).

19. Adeli, H. & Jiang, X. *Intelligent Infrastructure: Neural Networks, Wavelets, and Chaos Theory for Intelligent Transportation Systems and Smart Structures*. (CRC Press, 2008).

20. Fethi, M. D. & Pasiouras, F. Assessing Bank Performance with Operational Research and Artificial Intelligence Techniques: A Survey. *SSRN Electron. J.* (2009). doi:10.2139/ssrn.1350544

21. Nunes, I. & Jannach, D. A systematic review and taxonomy of explanations in decision support and recommender systems. *User Model. User-adapt Interact.* **27,** 393–444 (2017).

22. Makridakis, S. The forthcoming Artificial Intelligence (AI) revolution: Its impact on society and firms. *Futures* **90,** 46–60 (2017).

23. Bissio, R. Vector of hope, source of fear. *Spotlight Sustain. Dev.* 77–86 (2018).

24. Brynjolfsson, E. & McAfee, A. *The second machine age : work, progress, and prosperity in a time of brilliant technologies*.

25. Dobbs, R. *et al.* POORER THAN THEIR PARENTS? FLAT OR FALLING INCOMES IN ADVANCED ECONOMIES. *McKinsey Glob. Inst.* **7,** (2016).

26. Francescato, D. Globalization, artificial intelligence, social networks and political polarization: New challenges for community psychologists. *Community Psychol. Glob. Perspect.* **4,** 20–41 (2018).

27. Harrer, N. J. S. and A. Simulating norms, social inequality, and functional change in artificial societies.

28. Dalenberg, D. J. Preventing discrimination in the automated targeting of job advertisements. *Comput. Law Secur. Rev.* **34,** 615–627 (2018).

29. World Economic Forum (WEF). *Fourth Industrial Revolution for the Earth Series*

*Harnessing Artificial Intelligence for the Earth*. (2018).

30. Keramitsoglou, I., Cartalis, C. & Kiranoudis, C. T. Automatic identification of oil spills on satellite images. *Environ. Model. Softw.* **21,** 640–652 (2006).

31. Mohamadi, A., Heidarizadi, Z. & Nourollahi, H. Assessing the desertification trend using neural network classification and object-oriented techniques. *J. Fac. For. Istanbul Univ.* **66,** 683–690 (2016).

32. Maher, S. P., Randin, C. F., Guisan, A. & Drake, J. M. Pattern-recognition ecological niche models fit to presence-only and presence-absence data. *Methods Ecol. Evol.* **5,** 761–770 (2014).

33. De Fauw, J. *et al.* Clinically applicable deep learning for diagnosis and referral in retinal disease. *Nat. Med.* **24,** 1342–1350 (2018).

34. Russell, S., Dewey, D. & Tegmark, M. Research Priorities for Robust and Beneficial Artificial Intelligence. *AI Mag.* **34,** 105–114 (2015).

35. Gandhi, N., Armstrong, L. J. & Nandawadekar, M. Application of data mining techniques for predicting rice crop yield in semi-arid climatic zone of India. *2017 IEEE Technological Innovations in ICT for Agriculture and Rural Development (TIAR)* (2017). doi:10.1109/tiar.2017.8273697

36. Esteva, A. *et al.* Corrigendum: Dermatologist-level classification of skin cancer with deep neural networks. *Nature* **546,** 686 (2017).

37. Cao, Y., Li, Y., Coleman, S., Belatreche, A. & McGinnity, T. M. Detecting price manipulation in the financial market. *2014 IEEE Conference on Computational Intelligence for Financial Engineering & Economics (CIFEr)* (2014). doi:10.1109/cifer.2014.6924057

38. Beyer, H. L., Dujardin, Y., Watts, M. E. & Possingham, H. P. Solving conservation planning problems with integer linear programming. *Ecol. Model.* **328,** 14–22 (2016).

39. Whittaker, M. & et al. *AI Now Report 2018*. (2018).

40. West, J. & Bhattacharya, M. Intelligent financial fraud detection: A comprehensive review. *Comput. Secur.* **57,** 47–66 (2016).

41. Hajek, P. & Henriques, R. Mining corporate annual reports for intelligent detection of financial statement fraud – A comparative study of machine learning methods. *Knowledge-Based Syst.* **128,** 139–152 (2017).

42. Perry, W. L., McInnis, B., Price, C. C., Smith, S. C. & Hollywood, J. S. *Predictive Policing: The Role of Crime Forecasting in Law Enforcement Operations*.

43. Gorr, W., Neill, D. B. & Gorr, W. L. *Detecting and preventing emerging epidemics of crime*. (2007).

44. Ferguson, A. G. *The rise of big data policing : surveillance, race, and the future of law enforcement*.

45. IEEE. Ethically Aligned Design - Version II overview. (2018). doi:10.1109/MCS.2018.2810458

46. European Commission. Draft Ethics guidelines for trustworthy AI | Digital Single Market. (2018). Available at: https://ec.europa.eu/digital-single-market/en/news/draft-ethics-guidelines-trustworthy-ai. (Accessed: 30th January 2019)

47. Lipton, Z. C. The mythos of model interpretability. *Commun. ACM* **61,** 36–43 (2018).

48. Harari, Y. N. Yuval Noah Harari: the myth of freedom. *The Guardian* (2018). Available at: https://www.theguardian.com/books/2018/sep/14/yuval-noah-harari-the-new-threat-to-liberal-democracy. (Accessed: 30th January 2019)

49. Butler, A. J., Thomas, M. K. & Pintar, K. D. M. Systematic Review of Expert Elicitation Methods as a Tool for Source Attribution of Enteric Illness. *Foodborne Pathog. Dis.* **12,** 367–382 (2015).

50. Morgan, M. G. Use (and abuse) of expert elicitation in support of decision making for public policy. *Proc. Natl. Acad. Sci. U. S. A.* **111,** 7176–84 (2014).